\documentclass{imsart}

\RequirePackage{amsthm,amsmath}
\RequirePackage[numbers]{natbib}
\RequirePackage[colorlinks,citecolor=blue,urlcolor=blue]{hyperref}
\RequirePackage{graphicx}

\startlocaldefs

\RequirePackage{amssymb,bm}
\RequirePackage[mathscr]{eucal}
\DeclareFontEncoding{FMS}{}{}
\DeclareFontSubstitution{FMS}{futm}{m}{n}
\DeclareFontEncoding{FMX}{}{}
\DeclareFontSubstitution{FMX}{futm}{m}{n}
\DeclareSymbolFont{fouriersymbols}{FMS}{futm}{m}{n}
\DeclareSymbolFont{fourierlargesymbols}{FMX}{futm}{m}{n}
\DeclareMathDelimiter{\hsnorm}{\mathord}{fouriersymbols}{152}{fourierlargesymbols}{147}
\newcommand{\hs}{\big\hsnorm}

\newcommand{\tr}{{\rm tr}}

\usepackage{marginnote}

\endlocaldefs

\begin{document}

\begin{frontmatter}

\title{Tonal coarticulation revisited: functional covariance analysis to investigate the planning of co-articulated tones by Standard Chinese speakers}

\runtitle{Tonal coarticulation in mandarin by Standard Chinese speakers}

\begin{aug}

 \author{\fnms{Valentina} \snm{Masarotto}\ead[label=e1]{v.masarotto@math.leideuniv.nl}}
   \address{Mathematisch Instituut\\ Universiteit Leiden\\ Netherlands\\ \printead{e1}\\}

 \and

\author{\fnms{Yiya} \snm{Chen}\ead[label=e2]{yiya.chen@hum.leidenuniv.nl}}
   \address{Centre for Linguistics\\ and \\ Leiden Institute for Brain and Cognition \\Universiteit Leiden\\ Netherlands\\ \printead{e2}\\}

   \runauthor{V.~Masarotto and Y.~Chen}

 \end{aug}

\begin{abstract}
We aim to explain whether a stress memory task has a significant impact on tonal coarticulation. We contribute a novel approach to analyse tonal coarticulation in phonetics, where several f0 contours are compared with respect to their vibrations at higher resolution, something that in statistical terms is called variation of the second order. We identify speech recording frequency curves as functional observations and harness inspiration from the mathematical fields of functional data analysis and optimal transport. By leveraging results from these two disciplines, we make one key observation:
we identify the time and frequency covariance functions as crucial features for capturing the finer effects of tonal coarticulation. This observation leads us to propose a 2 steps approach where the mean functions are modelled via Generalized Additive Models, and the residuals of such models are investigated for any structure nested at covariance level. If such structure exist, we describe the variation manifested by the covariances through covariance principal component analysis. The 2-steps approach allows to uncover any variation not explained by generalized additive modelling, as well as fill a known shortcoming of these models into incorporating complex correlation structures in the data. The proposed method is illustrated on an articulatory dataset contrasting the pronunciation non-sensical bi-syllabic combinations in the presence of a short-memory challenge. 
\end{abstract}

\end{frontmatter}

\tableofcontents

\section{Introduction}
\label{sec:intro}
Most human form of communication comprise a melodic component. In European languages, the melodic component of speech is embodied in the variation of pitch. Volume
and intonation are used to capture attention, express feelings, differentiate affirmations from questions, and serious remarks
from sarcastic ones, making pitch variation an integral and essential part in conveying a message. Pitch variation
however can play an even more dramatic role when moving into the so called tonal languages. In tonal languages, the modulation of the pitch of the sound characterises the lexical identity of a word and conveys different lexical meanings. An intriguing aspect of tonal languages is tonal \textbf{coarticulation} (see e.g. \citet{recasens2018coarticulation}), which refers to the influence that the pitch or tone of one syllable has on the pitch or tone of adjacent syllables, leading to changes in how tones are realized in context. Indeed in everyday speech, words (and tones) are
rarely pronounced in isolation, but are always concatenated to one another. This implies that each tone is not stable, but is
affected by the neighbouring tones, preceding and following the tone of interest. The effect is primarily acoustic and perceptual.
Researcher in phonetics and experimental linguistic use various different ways to transform speech recordings into quantitative data (see, for example, \citet{cooke1993visual}). Most of them are based on the idea of representing sounds through 
\textbf{fundamental frequency curves}, denoted by f0. 
An f0 curve is a graphical representation of the fundamental frequency of a sound wave over time. The fundamental frequency is the lowest frequency of a sound wave and is perceived by the listener as the "pitch" of the sound. The amplitude of f0 quantifies the approximate frequency of the structure of spoken utterance
and measures objectively how high or low a speaker’s voice sounds. 
Tonal realization in connected speech is further influenced by more global higher-level factors (such as marking phrases, and sentence modes), which lead to  significantly deviated f0 contours of the voice from those produced in isolation (see review in \citet{chen2012oxford} for a more in depth description from a phonetical point of view). 

The existing body of research on local tonal coarticulation primarily focuses on investigating the directionality and magnitude of contextual effects on tonal production among native speakers, as evident in previous reviews such as those by \citet{chen2012oxford} and \citet{xu1997contextual}.  However, these studies have revealed diverse and occasionally contradictory patterns. As a whole, research on tonal coarticulation patterns and mechanisms remains scarce, making it a significant area of investigation in linguistics. 
This paper utilize data collected by \citet{zou2017production}, which aims to revisit tonal coarticulation with data from Standard Chinese. The data includes bi-syllabic speech recording from several native Mandarin speakers. A unique aspect of these recordings is the introduction of cognitive load, where speakers were asked to perform a task that increases their short-term memory while producing speech. The objective of the data collection is to support linguistic research on tonal coarticulation and help to adjudicate the current conflicting perspectives on tonal coarticulation patterns and the underlying cognitive mechanisms. Specifically, \textbf{our research questions aims to explain whether a stress memory task as cognitive load has a significant impact on coarticulation}.

Linguistic literature makes extensive use of sophisticated
statistical techniques to perform diverse analysis on f0 curves. The recent survey \citet{tavakoli2024statistics} gives a wide and wonderfully referenced overview of the most common areas of research in phonetics together with the most widely used analysis techniques. Typical analysis methods are growth curve analysis (\citet{mirman2017growth}), mixed models (\citet{winter2016analyze}) and more recently Generalised Additive Models (\citet{hastie1990generalized,wood2006generalized}). Generalized additive models (GAMs) and Generalized additive Mixed Models (GAMMs) are a class of statistical models that extend traditional linear models by allowing for non-linear relationships between the dependent and independent variables. 
This flexible modeling approach allows for complex, non-linear relationships to be captured in the data, and can improve the accuracy and interpretability of the output. GAMMs are not new, but only recent computational progress made
possible to analyse with them dataset often of millions of
data points in dynamic phonetic data. As a result, various recent
studies have seen the rise in the use of GAMMs, illustrating the potential of the
methods. \citet{meulman2015age} used GAMMs to
analyze EEG trajectories over time, and contextually showed how GAMMs method
were more sensitive than traditional ANOVA method. \citet{nixon2016temporal} performed a study on Cantonese tone
perception using GAMMs to analyse eye tracking data. \citet{wieling2016investigating} used GAMMs to compare Dutch dialect speakers. Finally, the
very hands-on tutorial by \citet{wieling2018analyzing}
made such methods widely understandable to
the linguistic community, and yet even more references can be found in the survey survey \citet{tavakoli2024statistics}, targeted both to a statistical and linguistic audience.

GAMMs are a powerful and flexible tool, but they also rely on a "discretized" version of the f0 curves, which are instead usually measured in Hz and most often assumed
to be smooth and continuous throughout their trajectories. The intrinsic smoothness of the f0 curves points naturally to the functional 
data analysis (FDA) domain (\citet{ramsay-silverman-2005, ferraty:2006}) as a tool to study them.

Accounting for the smooth structure of the underlying process helps in dealing with the high dimensionality of the data objects. Previous works which applied functional data analysis to linguistic
data included lip-motion \citet{ramsay1996functional}, analysis of temporal
prosodic effects \citet{byrd2006far} and speech
production \citet{koenig2008speech}. Recently however there seems to be a surge of successful application of FDA to linguistic, as it became apparent from the recent works \citet{tavakoli:2014, aston2010linguistic, pigoli2018statistical,koshy2022exploring,gubian2015using}, who use various aspect of functional data, from functional dimentionality reduction analysis in \citet{gubian2015using} to second order analysis via functional covariances in \citet{aston2010linguistic,pigoli2018statistical}. 

In this paper, we aim to apply some novel techniques of functional data analysis to complement the analysis of pitch variation and tonal coarticulation in Mandarin Chinese commonly performed via GAMMs. 
What most FDA applications have in common, with the exception of \citet{pigoli2018statistical}, is that they consider tonal variation of f0 curves only with respect to their shape, or their mean structure. However, functional data can exhibit another intriguing kind of variation, something that statistically is known as "second order variation". Second order variation is described by functional covariance operators,
and refers to the case where covariances are themselves the object of  statistical variation. These situations typically arise when
several covariance operators can be distinguished from data clusters. 
Recent literature
(\citet{aston2010linguistic,pigoli2018statistical}) offers evidence that significant
phonetic features are nested in the covariance structure between pitch
intensities at different frequencies. In fact, \citet{aston2010linguistic} claims
this covariance structure can be considered ``a summary of what a
language sounds like", while disregarding specific difference at
word level. Research in this direction is recently gaining traction, \citet{aston2010linguistic} employed functional
principal component to model pitch variation in phonetic data, 
\citet{hadjipantelis2015unifying} discussed a unified approach
to treat simultaneously amplitude and phase variation in Mandarin
Chinese, while \citet{pigoli2018statistical} stressed the
fundamental role played by covariance functions while analysing
acoustic phonetic data. 

The novelty of our work consists in proposing a unified approach to study tonal variation of f0 curves both with respect to their
shape, as well as to their variation of second order, and to apply it to a dataset specifically collected to study the effect of memory task on coarticulation. To the best of our knowledge, no study considered tonal coarticulation as a second-order effect. Effect of cognitive load on f0 contour is then explored via GAMMs. In the particular dataset considered, generalized additive models do not seem to provide a conclusive evidence on the cognitive load effect on coarticulation, in a way disproving empirical evidence by experts while at the same time suggesting that if such effect is significant, it might be found at finer resolution. 
Thus, we moved onto analysing the covariance structure of
the f0 residuals curves, obtained after the effect on the mean
of each tonal combination was modelled via the GAMM model. We employed the functional ANOVA permutation test by \citet{masarotto2022transportation}, which was showed to have state of the art performance, to show that indeed, the cognitive load effect is a higher resolution effect. 

When the null hypothesis of equality is rejected, it is natural as a second step to wish to describe the variation manifested by the covariances, or indeed obtain a parsimonious representation thereof. Functional PCA on covariance operators serves this purpose, and it helps quantify the effect of cognitive load on the spectrum of the covariance operators.

The article is structured as follows. Section \ref{sec:method} describes the data collection procedure and results based on \cite{zou2017production}. The statistical tools employed are described in Section \ref{sec:stat}. In particular, 
the procedure behind the first order analysis via GAMMs can be found in Section \ref{sec:gam}. Section \ref{sec:basic} lays out the necessary definitions in functional data analysis, Section \ref{sec:fANOVA} describes a functional test that allows to test the hypothesis of equality between covariance operators with very high power, and Section \ref{sec:pca} describes the procedure behind the only instances of tangent space principal component analysis known to investigate modes of variation among covariance operators. Results are collected in Section \ref{sec:numerical}.

\section{Data}\label{sec:method}
\subsection{Participants}\label{subsec:data}
Twelve native Mandarin speakers participated in the experiment with 3 three males and 9 nine females (age: M = 26.3, SD = 3.0). All are from the Northern part of China and speak standard Mandarin fluently. Four are native speakers of Beijing Mandarin, and the other eight speak Standard Chinese as their dominant language. The data are collected by \citet{zou2017production} as part of her Ph.D. thesis.

\subsection{Data collection and f0 extraction}\label{subsec:material and method}
The study is based on speech recording of balanced non-sensical tonal combination. The stimuli, as adopted from \citet{xu1997contextual}, are disyllabic sequences of /ma ma/ syllables with all possible tonal bi-combination of four Mandarin tones: high tone (T1), rising tone (T2), low tone (T3), and falling tone (T4). Among these 16 combinations, the T1T1 sequence is close to a the lexical item meaning "mother", the others are either nonsense sequences that unnatural in Mandarin (due to semantic/pragmatic oddity). In the following, we will refer to the 16 items as nonce words. The stimuli were selected as their segmental sequences with nasal onsets provide an optimal context for f0 tracking with ease of segmentation.  

All 16 possible tonal combinations were tested with four repetitions
in two conditions: no-cognitive-load and
high-cognitive-load condition. The cognitive-load conditions were set according to \citet{lavie2004load} and consisted of retention intervals of a short memory task. The
participants were recorded individually in the Leiden University
Phonetics Lab using E-prime (44.1 kHz, 16 bit) with a Sennheiser
MKH416T microphone. They were asked to read the sequences given in
pinyin, with instructions in Standard Chinese.
In each cognitive load trial, the reading task was preceded by six one-digit numbers to memorize, and followed by memory testing material. Participants were asked to try their best to remember the digits during the presentation of the memory material. 

The digits in the memory set were selected from 1 to 8 in random order, under the condition that the same digit never appeared more than twice in a trial, and no more than two digits appeared in sequential order. 

Each trial consisted of four repetitions of the 16 disyllabic
sequences, giving a total of 64 trials for both the cognitive load and the control setting. The order of these two settings was also randomized across participants. In total, there were 128 trials (16 disyllabic combinations, 4 four repetitions, 2 conditions) for each subject (S1-S12).
However, not all
repetitions were usable for all subjects, as some of them were
overloaded with noise. Table \ref{table:repetition} breaks down in
details how many repetitions were included for each subject. The two cognitive load conditions considered (no-cognitive-load and high-cognitive-load)
 are referred to as 'CL0' and 'CL6'
  respectively. As we
would like to filter out subject-specific way of speaking while
performing the analysis, we do not consider the missing recording to
be an issue. Another point to mention, is the overall noisiness of the data, necessitating special care in handling outliers. However, for the purposes of this study, covariances are computed across different speakers, repetitions, and tones, which helps to mitigate the impact of noise and makes the analysis more robust (as it becomes visible from a descriptive analysis in Table \ref{fig:speakers} in Appendix A).

\begin{table}[ht]
   \begin{minipage}{.49\textwidth}
\centering
\begin{tabular}{r|rrrr}
 \textbf{CL0} & \multicolumn{4}{c}{\textbf{repetition}}\\
  \hline
\textbf{speaker} & 1 & 2 & 3 & 4 \\ 
  \hline
  S1 & 16 & 16 & 16 & 16 \\ 
  S2 & 16 & 16 & 15 & 15 \\
  S3 & 16 & 15 & 15 & 15 \\
  S4 & 15 & 16 & 16 & 16 \\ 
  S5 & 16 & 16 & 16 & 16 \\
  S6 & 16 & 15 & 13 & 15 \\
  S7 & 16 & 16 & 16 & 15 \\ 
  S8 & 15 & 16 & 16 & 16 \\ 
  S9 & 16 & 15 & 16 & 15 \\ 
  S10 & 16 & 16 & 16 & 16 \\ 
  S11 & 16 & 14 & 14 & 15 \\ 
 S12 & 16 & 16 & 16 & 16 \\ 
   \hline
\end{tabular}
\end{minipage}%
\begin{minipage}{.49\textwidth}
\centering
\begin{tabular}{r|rrrr}
  \textbf{CL6} & \multicolumn{4}{c}{\textbf{repetition}}\\
  \hline
 \textbf{speaker} & 1 & 2 & 3 & 4 \\ 
  \hline

  S1 & 16 & 15 & 15 & 16 \\ 
  S2 & 12 & 13 & 16 & 16 \\
  S3 & 13 & 13 & 14 & 15 \\ 
  S4 & 14 & 14 & 13 & 16 \\ 
  S5 & 16 & 16 & 16 & 16 \\
   S6 & 14 & 14 & 12 & 14 \\ 
   S7 & 16 & 16 & 16 & 16 \\ 
  S8 & 13 & 16 & 16 & 16 \\
  S9 & 14 & 14 & 12 & 14 \\
   S10 & 16 & 15 & 15 & 13 \\ 
  S11 & 15 & 15 & 16 & 15 \\ 
  S12 & 14 & 15 & 16 & 15 \\
  \hline
\end{tabular}
\end{minipage}%
\caption{Repetitions per speaker in absence (left) and presence (right) of
  cognitive load.}
  \label{table:repetition}
\end{table}

\section{Statistical methods}\label{sec:stat}

\subsection{First order analysis via Generalized Additive Mixed Models}\label{sec:gam}
The coarticulatory effects on f0 curves were examined through a first-order analysis using generalized additive mixed models (GAMMs) implemented in the R programming language. The analysis was performed using the \texttt{mgcv} package in \texttt{R}, which is specifically designed for fitting GAMMs.

GAMMs model are particularly convenient to model dynamic phonetic data
for two reasons. Firstly, such models comprise of linear,
non-linear and random predictors, and allow to 
automatically determine the non-linear relationship between predictors
and dependent variables. Secondly, once the model is fitted,
predictors' effect can be analysed separately, facilitating the
interpretation of the results. The survey \cite{tavakoli2024statistics} offers a very complete overlook of the usage of GAMMs in phonetics and it provides a multitude of useful references. 
For the study of f0 contour, several fixed factors and random factors
were employed. We worked on the assumption that the mean bisyllabic $TiTj$ words
explained by the GAMM model differ
from one another, for every speaker, combination and repetition.
Moreover, we accept that each bi-syllabic entry might
include speaker- and repetition-specific variational patterns,
which are incorporated as random effects in the model. The inclusion of
these as random effects are justified as we want a general model that
filters factor of age, gender, and
emotional state at the moment of the recording. Statistical
significance of random effects is not tested, as it is beyond the
scope of the analysis \footnote{However, it would constitute an interesting follow-up, as linguistics literature is more and more tuned towards investigating individual variations}. Perhaps we could mention something along the line that it could be useful information to tap further into.Assuming that the random-effect structure is
given, fixed effects are compared and analysed separately for every
tonal combination. Tone contour and cognitive
load as a fixed constant term. Cognitive load and the interaction
between cognitive load and following or preceding tone are added as
non-linear terms over time. Finally, ordered factors were used to tap into potential cognitive load effect on particular tonal difference.

\subsection{Basic notions and notation for second order analysis}\label{sec:basic}
As stated in the introduction, we aim to complement the GAMM analysis and are interested in exploring whether cognitive load has an effect not (only)
on the mean structure of the curves (as captured by the GAMMs model)
but rather on the oscillation \emph{around} the mean, and thus be
identified as a so called \emph{second order} effect. The intuition behind the work is motivated by  \citet{aston2010linguistic}, where the authors demonstrate the covariance structure between the intensities at different frequencies captures language-specific variability (how a language "sounds like") and it does not incorporate
differences at the word level.
In particular, in case the GAMM analysis does provide an acceptable fitting but not significant cognitive load insight, we would like to check whether the effect of
cognitive load can be nested into the covariance structure of the residuals of the GAMM
model. 

In phonetics, a $f0$ curve is a graphical representation of the fundamental frequency of a sound wave, and it is generally regarded as a smooth function of frequency over time. Since we can represent each spoken word
as smooth function, we turn to functional data analysis (see also \cite{tavakoli2024statistics} for FDA references in phonetics). 
Formally, we call a functional observation a random element $X$ defined over a probability space $(\Omega, \mathcal{F},\mathbb{P})$ and normally belonging to a real separable (possibly infinite dimensional) Hilbert space $\mathcal{H}$, equipped with the inner product $\langle\cdot,\cdot\rangle:\mathcal{H}\times\mathcal{H}\rightarrow\mathbb{R}$, and corresponding norm $\|\cdot\|:\mathcal{H}\to[0,\infty)$. Here, unless differently specified,  we take $\mathcal{H}$ to be a bounded time domain, such as $X\in L^2[0,T]$ with $\mathbb{E}(||X||^2)<\infty$. 
In this specific work, we are interested in covariance operators. Covariances are mathematical operator that describe the relationship between the variations in two different argument values. The covariance operator is defined as 
$$\Sigma=\mathbb{E}\left[(X-m)\otimes (X-m)\right]:=\int_{\Omega}(X-m)\otimes(X-m)d\mathbb{P},$$
where $m$ is the mean function
$$m=\mathbb{E}(X):=\int_\Omega X d\mathbb{P}.$$
We work in the scenario of $N$ independent samples $\{X_{1,j}\}_{j=1}^{n_1}, \dots ,\{X_{N,j}\}_{j=1}^{n_N}$ of i.i.d. random elements in $\mathcal{H}$. Each of the $N$ sample is modelled by a prototypical random function $X_i$, with mean function $\mu_i\in \mathcal{H}$ and covariance
operator $\Sigma_i:\mathcal{H}\times \mathcal{H}\to\mathcal{H}$ and we observe $n_i$ realisations from each population.
In applications one inevitably works with  finite-dimensional
representations of the mean and the covariance functions, which can be seen as functional analogues of the mean vector and variance-covariance matrix in multivariate data analysis. For the rest of the paper, when referring to the mean or the covariance operators, we will actually implicitly refer to their sample version, defined by

\begin{align*}
\widehat\mu_i &= \dfrac{1}{n} \sum_{j=1}^n X_{i,j}\\
\widehat\Sigma_i &= \dfrac{1}{n-1} 
\sum_{j=1}^n \bigl(X_{i,j}-\widehat\mu_i\bigr)\otimes\bigl(X_{i,j}-\widehat\mu_i\bigr).
\end{align*}

We refer to \citet{masarotto2018procrustes} for a mathematical justification as to why functional results employed in this paper do hold for the finite dimensional projection above.

\subsection{Functional ANOVA to compare unexplained variation with and
  without cognitive load}\label{sec:fANOVA}
A common assumption in linguistic is that covariance operators are common to all the words
within each language. In our case, it translates into the covariance operators being common to ll replications of
the same two syllables, once the curves have been aligned with respect to their means. This assumption is compatible with recent works such as \citet{aston2010linguistic,pigoli2018statistical} where the authors show that the covariance structure of $f0$ curves is language-specific and not word-specific. However, these works do not specifically consider tonal languages. We want to lay out the necessary theory to check the validity of this assumption in Mandarin Chinese. We claim that we can identify an effect of cognitive load on coarticulation, in a way identifying tonal characterization as a bearer of differences as isolated as different roman languages in \citet{aston2010linguistic}.
We perform the check by considering a suitable, highly sensitive distance between two estimates, and performing a distance-based permutation test. One might wonder whether distances between two finite-dimensional representations provides a good approximation
of the corresponding distance between the infinite-dimensional objects. As expected, this is the case, as proved in \citet{masarotto2022transportation}, and the same
can be said about transport maps that "deforms" one operator into another in equation \ref{maps}. The tests follows the functional ANOVA testing
procedure of \citet{masarotto2022transportation}, whose main intuition is summarized hereafter for
completeness.

Assume we have $N$ groups of functional data, $\{X_{i,j}\}$,
$i=1,\dots,n$, $j=1,\dots,N$.  Mark as $\Sigma_{i}$ the (empirical) covariances
relative to each group, $i=i,\dots,N$. \citet{masarotto2022transportation} propose a distance-based permutation test built around a test-statistics which exploits
elements of optimal transport, in particular the so-called optimal
transport maps. Optimal transport is a mathematical field that studies how to optimally deform one distribution of mass to another, while minimizing the cost of such deformation. The optimal deformation is obtained through optimal transport maps functions. The testing procedure rely on the notion of Fr\'echet mean. ${\overline\Sigma}$ is a Fr\'echet mean of the operators $\{\Sigma_i\}_{i=1}^{N}$ if
$$\overline\Sigma=\underset{\Gamma\in G}{\arg\min}\sum_{i=1}^{K}\textrm{dist}^2(\Sigma_i,\Gamma).$$
Fr\'echet means generalize arithmetic means to general metric spaces. Here $G$ is a suitable parametric function space and we take the distance to be the Wasserstein-Procrustes distance of optimal transport between covariance operators, whose expression (both in finite and infinite dimension) is 
$$
\Pi^2(\Sigma_1,\Sigma_2) = 
\inf_{\begin{smallmatrix} Z_1\sim N(0,\Sigma_1)\\Z_2 \sim N(0, \Sigma_2)\end{smallmatrix}} \mathbb{E}\|Z_1-Z_2\|^2 \\
     =\tr\Sigma_1+\tr\Sigma_2-2\tr\sqrt{\Sigma_1^{1/2}\Sigma_2\Sigma_1^{1/2}}.$$
     In practice, we are required to compute the unique empirical Fr\'echet mean $\hat\Sigma$,
\begin{equation}\label{eq:empiricalfrechet}
\hat\Sigma=\underset{\mathbb{R}^{q\times q}\ni\Gamma\succeq 0}{\arg\min}\sum_{j=1}^{K}\Pi^2(\hat\Sigma_j,\Gamma).
\end{equation}
For general cases, equation (\ref{eq:empiricalfrechet}) has no closed form formula, but
it can be approximated by the iterative steepest descent procedure described in
\cite[Section~8]{masarotto2018procrustes}.

We can now define the test statistic
\[
T=\sum_{i=1}^{N}\hs \mathbf{t}_i-\mathscr{I}\hs_2^2,
\]
where $\mathscr{I}$ is the identity operator, $\hs \cdot \hs_2$ denotes the Hilbert-Schidt norm, and
\begin{equation}\label{maps}
    \mathbf{t}_i={\overline\Sigma}^{-1/2}({\overline\Sigma}^{1/2}\Sigma_j{\overline\Sigma}^{1/2})^{1/2}{\overline\Sigma}^{-1/2},\quad i=1,\ldots,N
\end{equation}
are optimal transport maps. Again, in practice, we will work work with the empirical versions of the $\mathbf{t}_i$. In \cite{masarotto2022transportation}, the transport-maps based test in showed to be state of the art, and extremely sensitive in case of high-frequency functional differences.  

\subsection{Functional Principal Component Analysis of Covariances}\label{sec:pca}
If the null
hypothesis of equality among covariances is rejected, a second
order analysis would entail explaining and understanding the variability of
the sample of covariances around its Fr\'echet mean, and possibly interpreting the main
directions of this variation. 
As in multivariate analysis in Euclidean spaces, Principal Component Analysis (PCA) is a
good candidate for such task. When it comes to covariances, PCA is hindered by their non-linearity. However, \citet{masarotto2022transportation} solved this obstacle presenting the first instance of tangent space PCA of covariances, which we summarize hereafter. 

Once the empirical Fr\'echet mean $\widehat\Sigma$ of the
operators $\widehat{\Sigma}_i$ is available, we can perform functional PCA by suitably mapping them onto a linear (tangent) space. We follow the steps of \citet{masarotto2022transportation} carry out tangent space PCA by performing regular PCA through the spectral decomposition of the empirical operator
  $$\frac{1}{K}\sum_{j=1}^K \left[(\hat{\mathbf{t}}_j-\mathscr{I})\hat\Sigma^{1/2}\right]\otimes\left[
  (\hat{\mathbf{t}}_j-\mathscr{I})\hat\Sigma^{1/2}\right]$$
and use the result to investigate the
  variability of the dataset.

\section{Results}\label{sec:numerical}
This Section presents
the first and second order analysis
the interaction effects of tonal context and cognitive load. 
In particular, functional ANOVA allows to detect a significant difference between the covariance operators relative to curves with
cognitive load and without cognitive load, as presented in Section \ref{sec:fun_ANOVA_results}. Covariance principal
component is used to tap into such difference in Section \ref{sec:pca_results}.

\subsection{First order analysis via Generalized Additive Models}
\begin{figure}[ht]
\centering 
    \includegraphics[width=\textwidth,height=12cm]{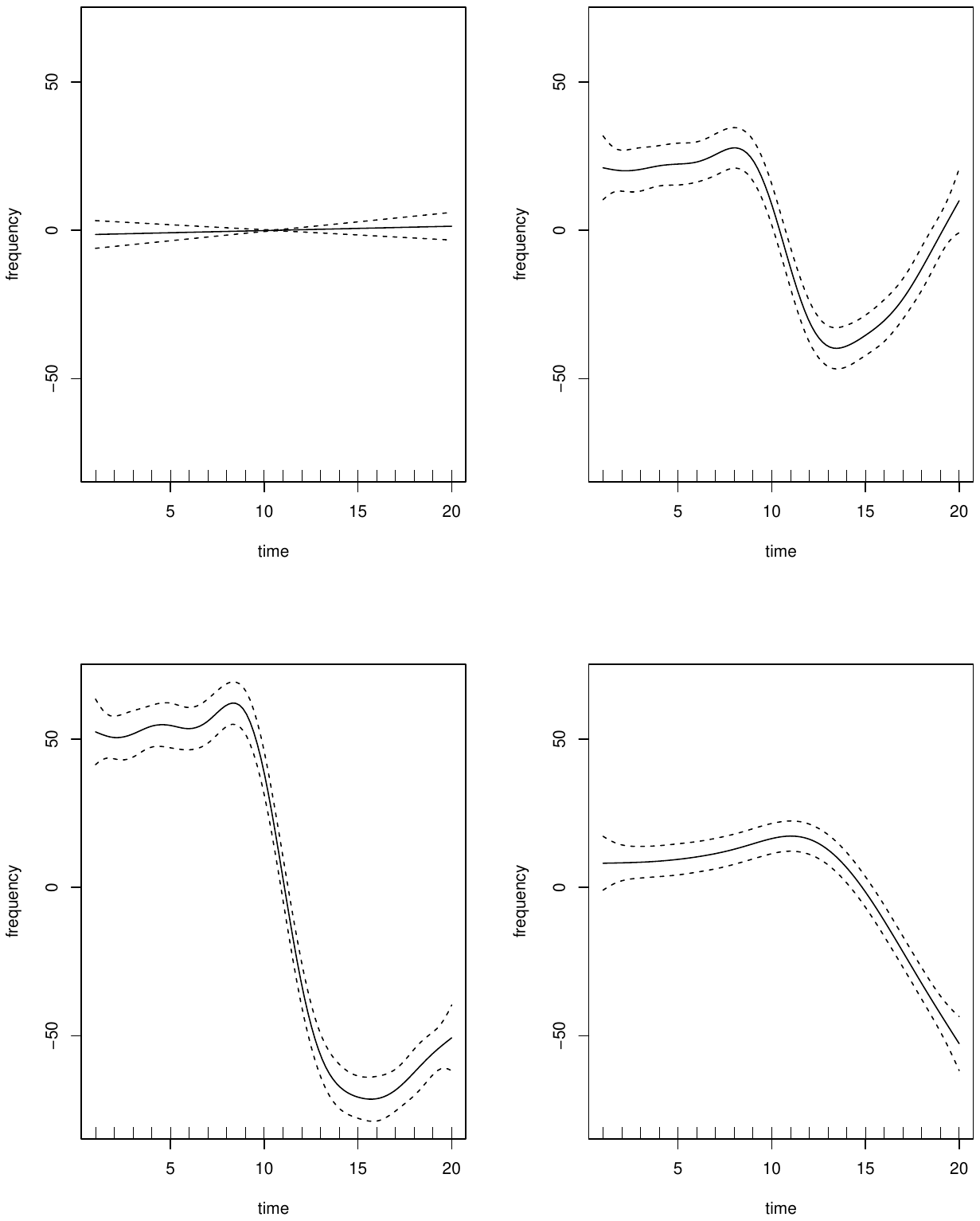}
    \caption{Average across repetitions and speakers of T1Tx f0 contours}
    \label{fig:t1tx}
\end{figure}

\begin{figure}[ht]
\centering 
    \includegraphics[width=\textwidth,height=12cm]{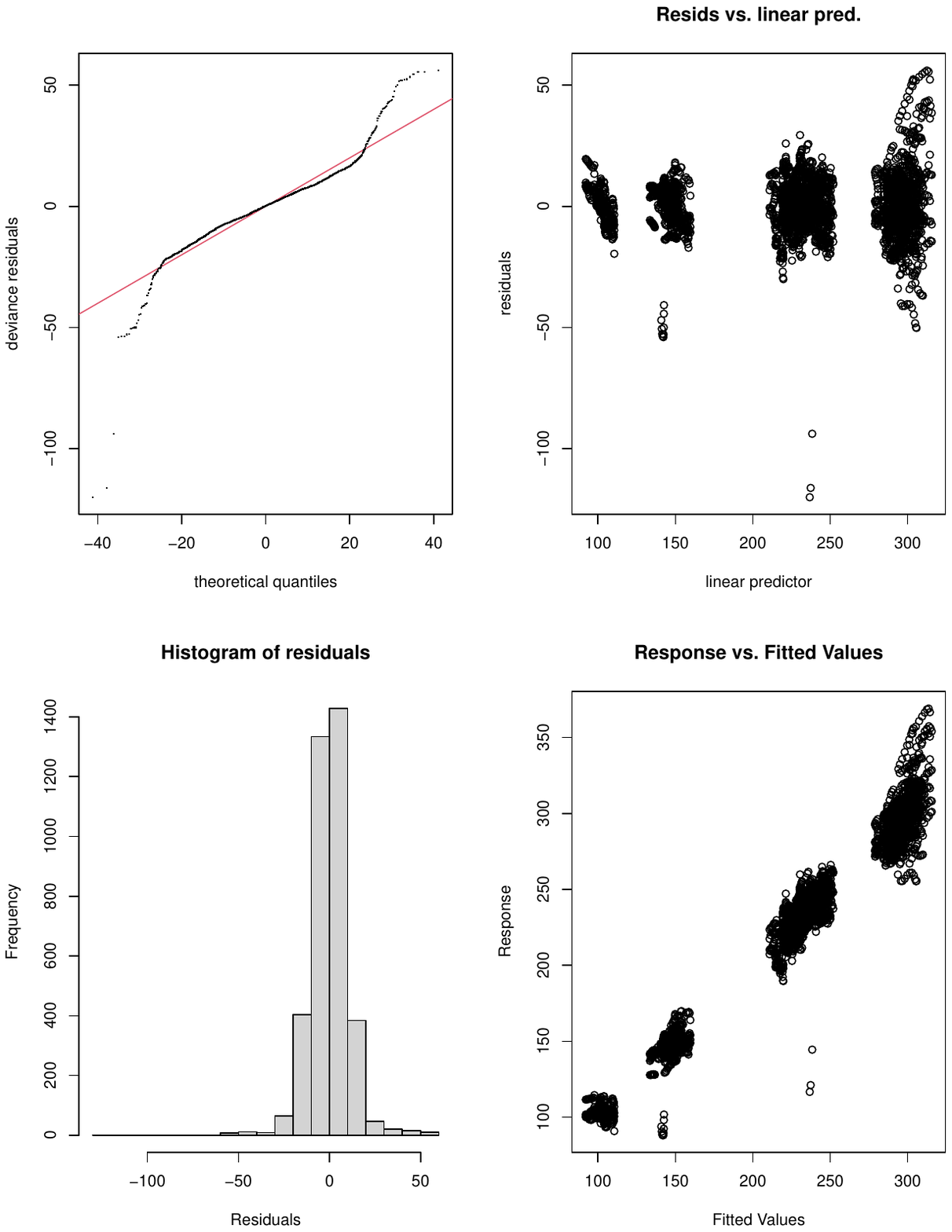}
    \caption{GAM model check for T1Tx F0 contours}
    \label{fig:gamT1Tx}
\end{figure}

GAMMs were employed to assess coarticulatory effects on f0
contour. The analysis were done via the \texttt{mgcv} package in R (version 1.8-23; \citet{wood2006generalized,wood2015package}).
Figure \ref{fig:t1tx} shows the average contour of the 4 tonal combinations considered. For better readability we include here the modelling of T1 in syllable 1 as a function of the following tones (i.e. anticipatory tonal coarticulation effect) and cognitive load. Other comparisons can be found in Appendix~\ref{appendix:gam}. 
Some speakers, or tone combination, will on average have a more pitch variation than others, and this structural variability is captured by 
random speaker intercept and slope. Simulations showed that it is not necessary to include differentiation by preceding or following syllable. An AR(1) error model is incorporated for residuals (assuming a Gaussian model). Smooths are added to model cognitive load, with interaction between cognitive load and following tone. Results are given in Table~\ref{table:gamT1Tx}. Figure~\ref{fig:gamT1Tx} gives a graphical check of the model. We can see that cognitive load has not at all or very little effect on tones T2 and T4, and only seems slightly detectable on the parametric coefficients. Similar results can be seen for other tonal combinations. In the next two Sections, we will investigate if a cognitive load effect exist but it is hidden at a higher resolution. 

\begin{table}
  \centering
  \begin{tabular}{|ll||ll|}
    \hline
 \textbf{Parametric coefficients} & \textbf{p-value} &\textbf{Smooth terms} &  
                                                             \textbf{p-value}\\
    \hline
    (Intercept) &  $<0.01^{***}$ & s(time): no CL T1.T1  &  $0.45$\\
    CL T1.T1 &  $<0.01^{***}$ & s(time): CL T1.T1 &  $0.69$\\
    CL T1.T2 & $0.20$ & s(time): CL T1.T2   & $0.04^{*}$\\
    CL T1.T3  & $<0.01^{***}$ & s(time): CL T1.T3  & $0.03^{*}$\\
    CL T1.T4  & $0.03^{*}$ & s(time): CL T1.T4  & $0.39$\\
    && s(time, subject interaction)  & $<0.01^{***}$\\
    \hline
  \end{tabular}
\caption{GAM output to model anticipatory tonal coarticulation effect in T1Tx combination.}
\label{table:gamT1Tx}
\end{table}

\subsection{Functional ANOVA of covariances}\label{sec:fun_ANOVA_results}

Let $X_{i,j,S_n,CL_l}$, $i,j=1,\dots,4$, $n=1,\dots,12$, $l\in{0,6}$ be the residuals curves of the GAMM model. We indicatee as $CL_0$ the lack of cognitive load (as to 0 ciphers to remember) and $CL_6$ the presence of cognitive load (as to 6 ciphers to remember). We fit a separate model for each TiTx and for each TxTj combination, $i,j=1,\dots,4$, yielding total of eighth models each giving rise to 384 residual curves (12 speakers, 4 repetitions, 4 tonal combinations where either the first or the second tone is fixed and two cognitive load conditions). In practice, not all these 384 curves are viable, as explained in Section~\ref{subsec:data}.
Let $\widehat{\Sigma}^{CL0}_{TiTj},\widehat{\Sigma}_{TiTj}^{CL6}$
$i,j=1,\dots,4$ be the bi-syllable specific empirical covariances extracted from the residuals of the GAMM model. 
We employ the permutation-based test defined in Section \ref{sec:fANOVA} to verify whether cognitive load has a second order effect. Intuitively, we want to verify whether can detect a significant difference induced by cognitive load in the covariance structure of the GAMMs residuals. In mathematical terms this translates into testing for every tonal combination $TiTj$, $i,j=1,\dots,4$ the null hypothesis of equality:
$$H_0: \widehat{\Sigma}^{CL0}_{TiTj}=\widehat{\Sigma}_{TiTj}^{CL6}$$
versus the alternative hypothesis that the two operators differ. 
To carry the test in practice, we generate $2n$ curves ($n$ for each cognitive load scenario) and unless otherwise stated, the sample size of each group is
$n=30$. The power of the test is estimated from 500 replications. The
number of permutations is 500. The procedure is repeated for each of
the 16 $TiTj$ tonal combination. Table \ref{table:fanova} collects the mean
p-values over 500 replications for each tonal combination. A significant p-value (less than 0.05)
would indicate that there is a significance difference between the
covariance structure of the residuals of the GAMMs model, when we
compare curves with and without cognitive load. 
It is evident from the
table that the covariance structure of the residuals carries a
significance difference between the curves with cognitive load, and
the curves without. The effect of cognitive load is independent on the
tonal combination considered. Given that the cognitive load seems to
be a second order effect, it is not surprising that the GAMMs
analysis, albeit correct, was only partially successful in modelling
it. After performing the global test, in
case the null hypothesis $H_0$ is rejected, one can investigate
main directions of variation through a tangent space
principal component analysis, as described in
\citet{masarotto2022transportation}. One could also look at pairwise differences with post-analysis comparison, as in
\citet{cabassi2017permutation,pesarin2010permutation}. We discuss this in the next paragraph. 
\begin{table}
  \centering
  \begin{tabular}[ht]{c|cccccc}
    & Min. & 1st Qu. &  Median & Mean & 3rd Qu. & Max. \\
    \hline
    T1T1 & 0.009 & 0.019 & 0.019 & 0.024 & 0.029 & 0.218 \\
    T1T2 & 0.009 & 0.019 & 0.029 & 0.052 & 0.079 & 0.257 \\
    T1T3 & 0.009 & 0.009 & 0.019 & 0.034 & 0.019 & 0.217 \\
    T1T4 & 0.009 & 0.009 & 0.009 & 0.016 & 0.049 & 0.039\\
    \hline
    T2T1 & 0.009 & 0.009 & 0.019 & 0.044 & 0.039 & 0.415\\
    T2T2 & 0.009 & 0.019 & 0.044 & 0.063 & 0.089 & 0.316\\
    T2T3 & 0.009 & 0.009 & 0.019 & 0.031 & 0.039 & 0.178\\
    T2T4 & 0.009 & 0.009 & 0.009 & 0.017 & 0.019 & 0.059\\
    \hline
    T3T1 & 0.009 & 0.009 & 0.009 & 0.011 & 0.009 & 0.029\\
    T3T2 & 0.009 & 0.009 & 0.009 & 0.011 & 0.009 & 0.029\\
    T3T3 & 0.009 & 0.009 & 0.009 & 0.010 & 0.009 & 0.029\\
    T3T4 & 0.009 & 0.009 & 0.009 & 0.012 & 0.009 & 0.039\\
    \hline
    T4T1 & 0.009 & 0.009 & 0.009 & 0.011 & 0.019 & 0.049\\
    T4T2 & 0.009 & 0.009 &  0.009 &  0.009 &  0.009 &  0.009\\
    T4T3 & 0.009 & 0.009 & 0.019 & 0.063 & 0.069 & 0.069\\
    T4T4 & 0.009 & 0.009 & 0.009 & 0.013 & 0.009 & 0.039\\
  \end{tabular}
\caption{$p$-values for the functional ANOVA test applied on the residual covariance
  operators. A significant $p$-value indicates a second-order effect of the cognitive load.}
  \label{table:fanova}
\end{table}

\subsection{Principal component analysis to understand modes of variation}\label{sec:pca_results}
\begin{figure}[ht]
\centering 
\includegraphics[width=\textwidth,height=10cm]{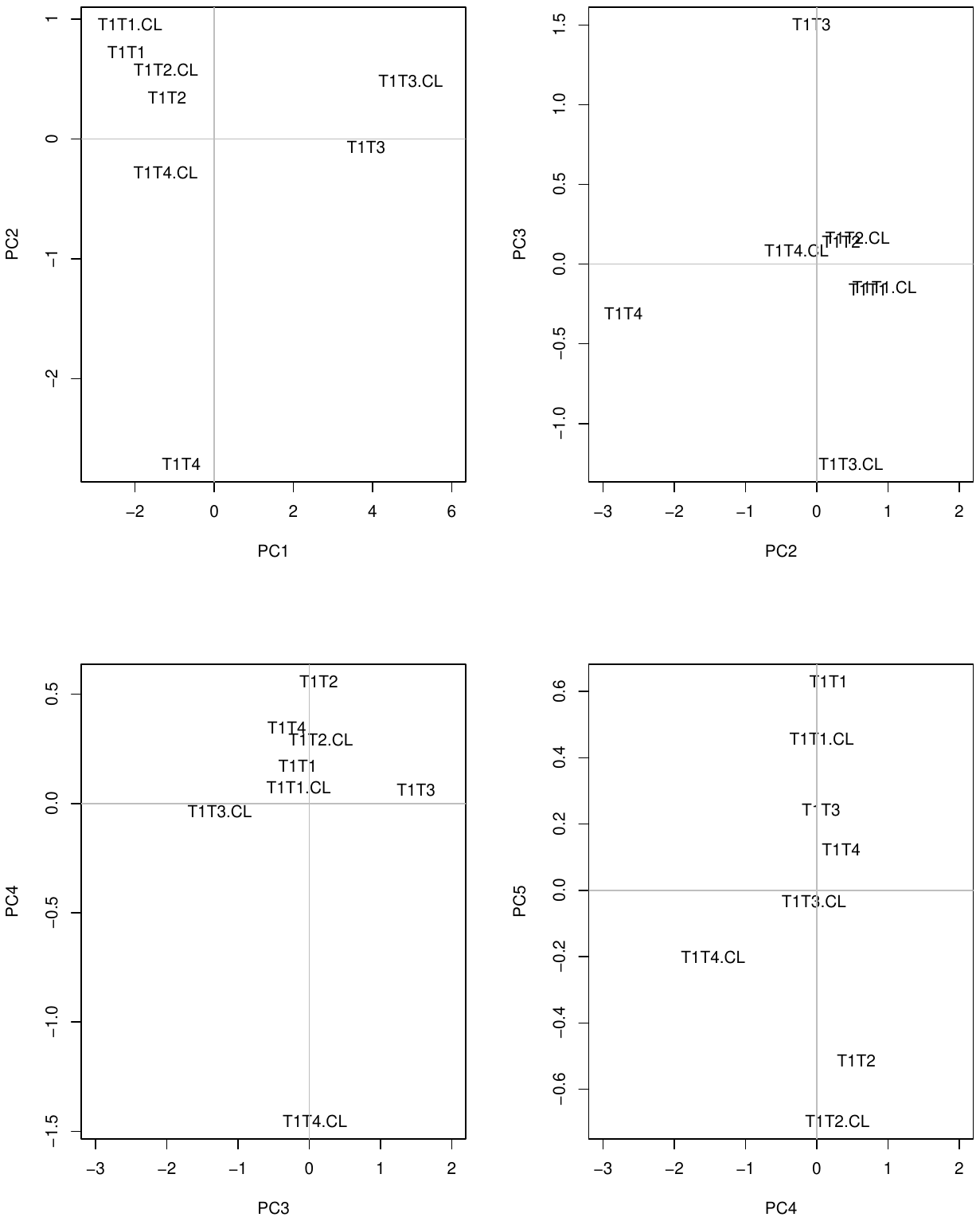}
\caption{PCA plot for T1Tj combination}
\label{fig:pcaT1Tx}
\end{figure}

A separate analysis is performed for all tonal combination beginning or ending with the
same tone. To slim the notation, we focus on the collection of curves corresponding to each tonal
combination $TiTx$, $i=1, \dots, 4$, which gives rise to two sample
covariance operators, one corresponding to $CL0$ and another one
corresponding to $CL6$. Figure \ref{fig:pcaT1Tx} shows the results of applying PCA on the tonal
combination $T1Tj$, $j=1,\dots,4$. We see clearly that tangent space PCA captures
very well the difference between cognitive load and lack thereof, as
most scenarios are isolated in at least one plot. More precisely:
 \begin{enumerate}
  \item The first PC captures (part of) the difference between T1T3 (with and without cognitive load) and all the other tonal combinations. This is to be expected, as T3 tone is realized with lot of glottalization, or creaky voice, and contributes the biggest fluctuation change in the f0 contours, thus moving from any tone to T3 requires a lot of amplitude variation, which is exactly what covariance operators capture. 
  \item   The second PC captures (part of) the difference between the tones T1T4 (without cognitive load) and all the other tones. 
  \item   The third PC captures (part of) the difference between the tones T1T3 with or without cognitive load. 
  \item   The fourth PC captures (part of) the difference between the tones T1T4 with or without cognitive load. 
  \item The fifth PC captures (part of) the differences between T1T1 and T1T2 tonal combination. Cognitive load effect cannot be separated in this case. 
  \item    Since, the order in the $y$-axes is the order of magnitude
    of the eigenvalues the analysis suggests also the importance of
    the differences between the operators. As intuition dictates the
    difference between the effect of cognitive load on T1T3 is nearly twice more pronounced than the effect of cognitive load on T1T4. 
\end{enumerate}

\begin{figure}[ht]
\centering 
\includegraphics[height=5cm]{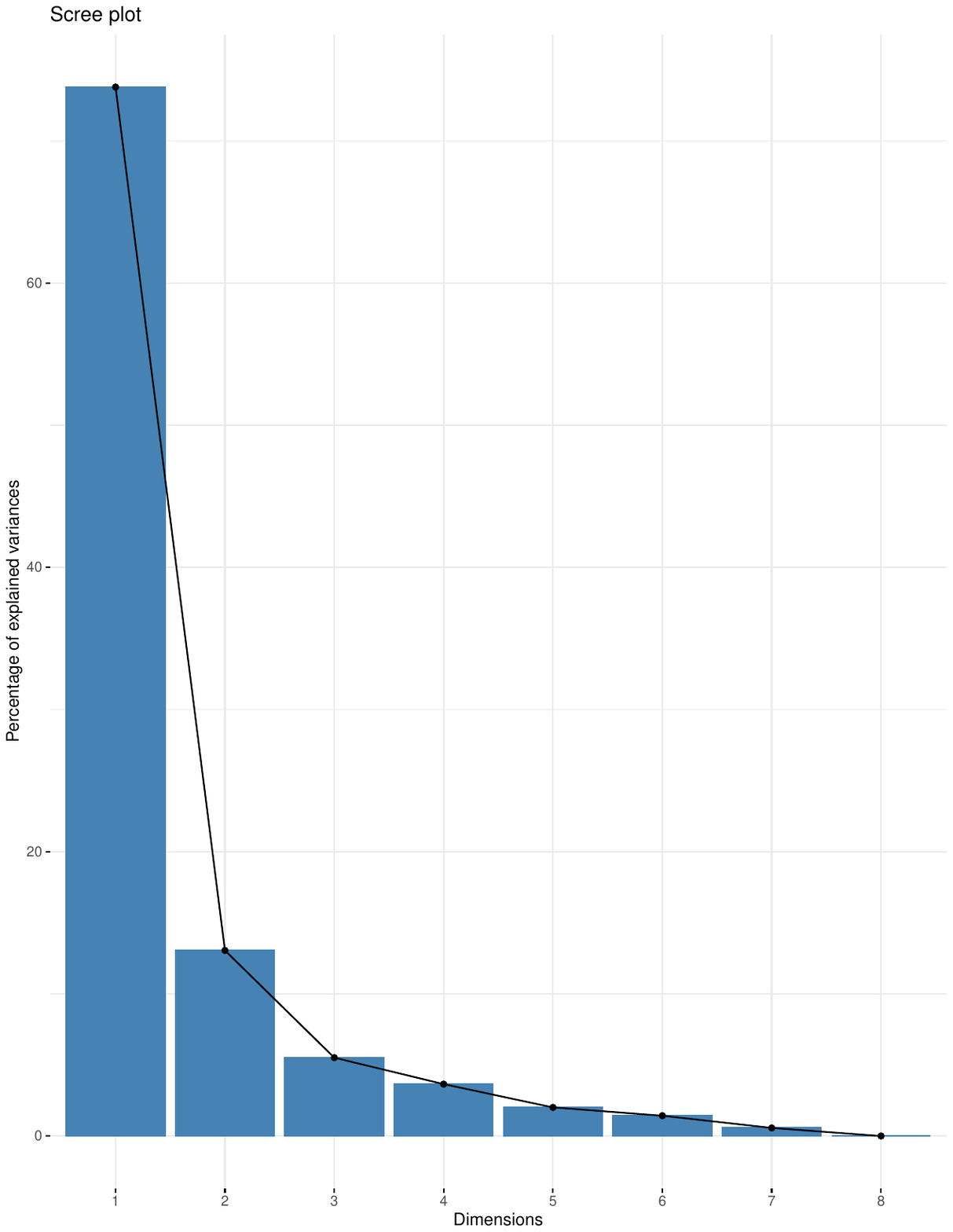}
\caption{Eigenvalues screeplot for T1Tj combination}\label{screeplotT1Tx}
\end{figure}

The screeplot (Figure~\ref{screeplotT1Tx}) shows that the first 3 PCs explain almost
the full variance of the data, but yet marking cognitive load as a
subtle effect. The fourth PC explains around 5$\%$ of the variance. A natural question while doing PCA is the interpretation of the effect of each PCs individually. A task which is in general not trivial, and which makes particularly pleasing the fact that we can interpret them in terms of cognitive load in this setting. 

\section{Discussions}
We have introduced a novel approach to investigate the effect of coarticulation that combines the usability of generalized additive (mixed-effects) modeling (GAMMs) with the power of FDA. We analyzed an articulatory dataset contrasting the pronunciation non-sensical bi-syllabic combinations in the presence of a memory challenge (cognitive load) in order to investigate the effect of such cognitive load onto tonal coarticulation. We could show that while cognitive load seems to have an effect on tonal coarticulation, such effect seems hidden at very fine resolution. 

With respect to the actual modeling, we have used the R packages mgcv \citet{wood2015package} and fdWasserstein \citet{masarotto2024package}. By using generalized additive modeling, we were able to analyze all dynamic data, and did not have to average over time or select a specific time point. The analysis allowed us to assess specific non-linear patterns, while simultaneously taking all dependencies in our data into account. While the generalized additive modeling approach is flexible and powerful, a known shortcoming is that no correlation structure can be incorporated in the linear random effects structure. The 2-steps analysis presented here allows to overcome such limitation. Analysing the residuals of the GAMM model revealed a non-captured effect of cognitive load on tonal coarticulation. Tangent space principal component analysis helped to visualize such effect. 

We would like to emphasize that more follow-up analysis are possible, for example, once an effect of cognitive load at covariance level is detected, pairwise comparison to tap into this effect in the fashion of \citet{pesarin2010permutation} is also possible. Moreover, one could investigate the effect of cognitive load on vowel duration via a Cox regression model (\citet{cox1972regression}). In this setting, time fixed covariates include the presence of cognitive
load, the different tone and the interactions between such terms. We leave the details to future work.

\section{Acknowledgments}
We would like to kindly thank Ting Zou for providing the data collected during her Ph.D. thesis (\citet{zou2017production}).

\bibliographystyle{imsart-nameyear}
\bibliography{biblio}

\appendix
\section{Difference across repetitions}
Table \ref{fig:speakers} displays the level of noise in the data for repetition 1. Similar differences are present in all repetitions. 

\begin{table}
\begin{tabular}{l}
\includegraphics[width=12cm,height=9cm]{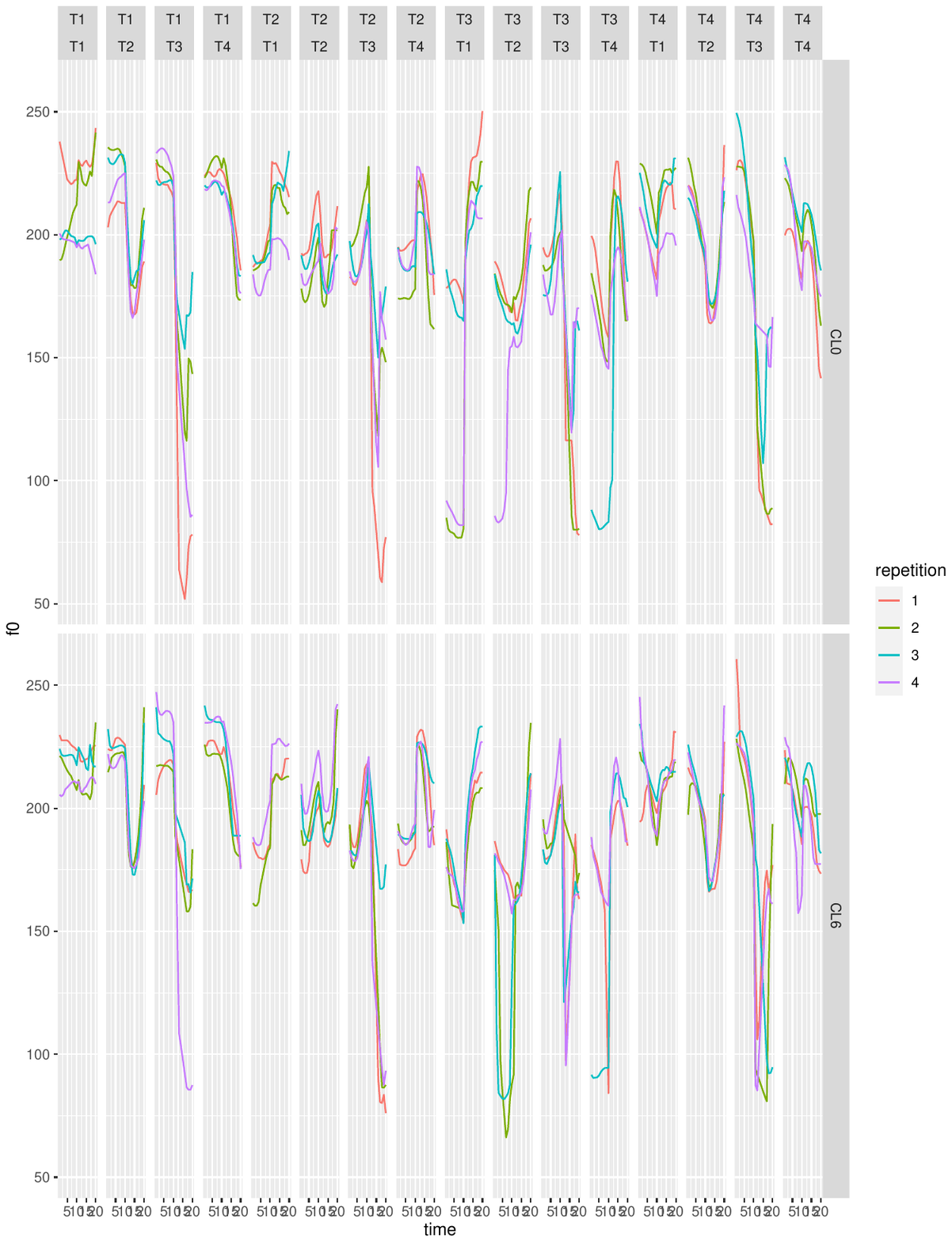} \\ \includegraphics[width=12cm, height=9cm]{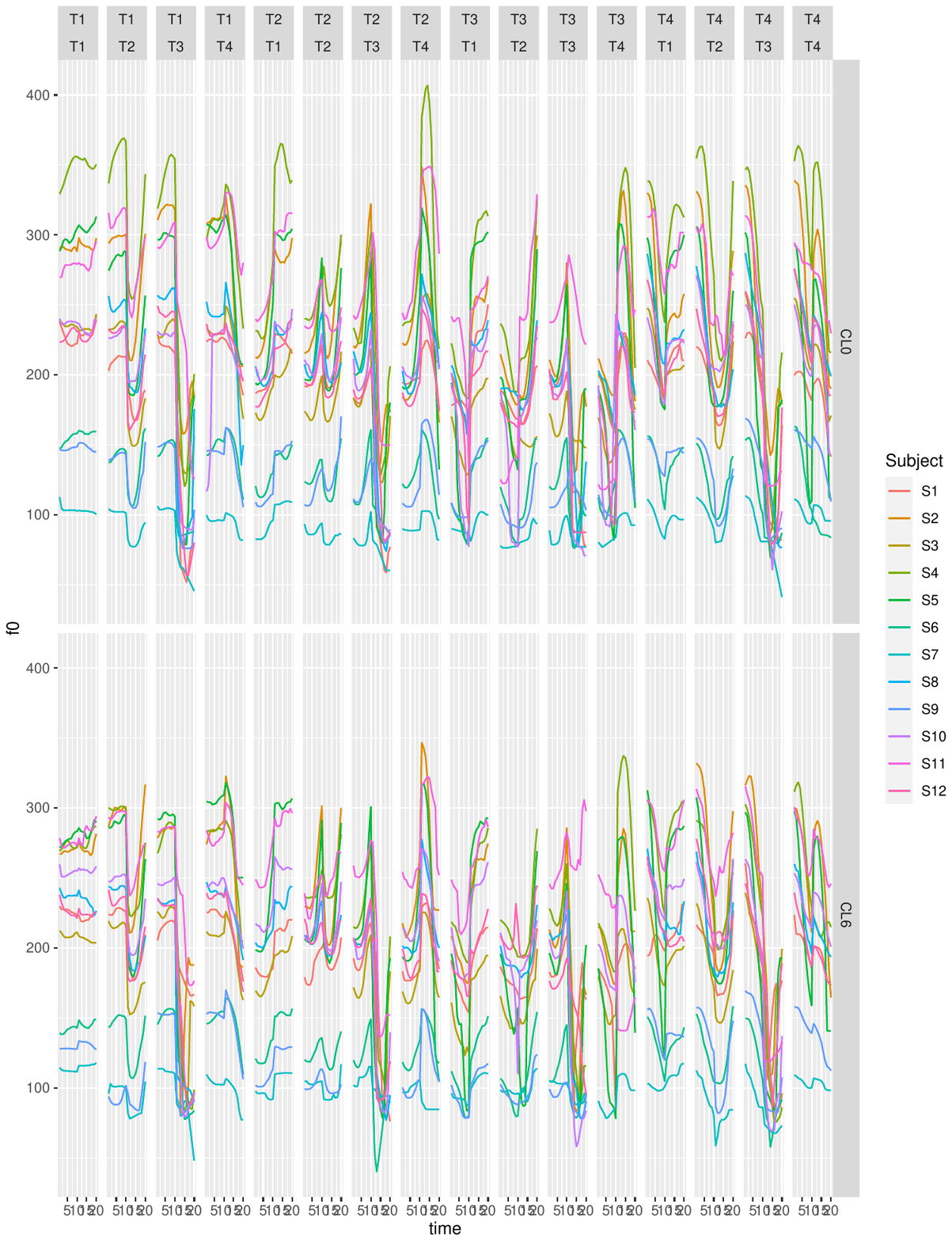}
\end{tabular}
\caption{Speaker 1 differences across repetitions (left) and f0 differences per repetition (right, first repetition plotted).}
\label{fig:speakers}
\end{table}

\section{f0 contours via GAM}\label{appendix:gam}
This Appendix contains the analysis the tonal combination not present in the main body. 
\newpage
\subsection{GAM model for T2Tx combination}
\vspace{-0.5cm}
\begin{table}[h!]
  \label{table:gamT2Tx}
  \centering
  \begin{tabular}{|ll||ll|}
    \hline
 \textbf{Parametric coefficients} & \textbf{p-value} &\textbf{Smooth terms} &  
                                                             \textbf{p-value}\\
    \hline
    (Intercept) &  $<0.01^{***}$ & s(time): no CL T2.T1  &  $0.01^{***}$\\
    CL T2.T1 &  $<0.07$ & s(time): CL T2.T1 &  $0.01^{***}$\\
    no CL T2.T2  & $<0.01^{***}$ & s(time): no CL T2.T2  & $0.01^{***}$\\
    CL T2.T2 & $0.01^{***}$ & s(time): CL T2.T2   & $0.01^{***}$\\
    no CL T2.T3  & $<0.01^{***}$ & s(time): no CL T2.T3   & $0.01^{***}$\\
    CL T2.T3  & $<0.01^{***}$ & s(time): CL T2.T3  & $0.01^{***}$\\
    no CL T2.T4  & $<0.01^{***}$ & s(time): no CL T2.T4  & $<0.01^{***}$\\
    CL T2.T4  & $0.62$ & s(time): CL T2.T4  & $0.01^{***}$\\
    && s(time, subject interaction)  & $<0.01^{***}$\\
    \hline
  \end{tabular}
\caption{GAM output to model anticipatory tonal coarticulation effect in T2Tx combination.}
\end{table}

\begin{table}[h!]
\begin{tabular}{l}
    \includegraphics[width=12cm, height=5.5cm]{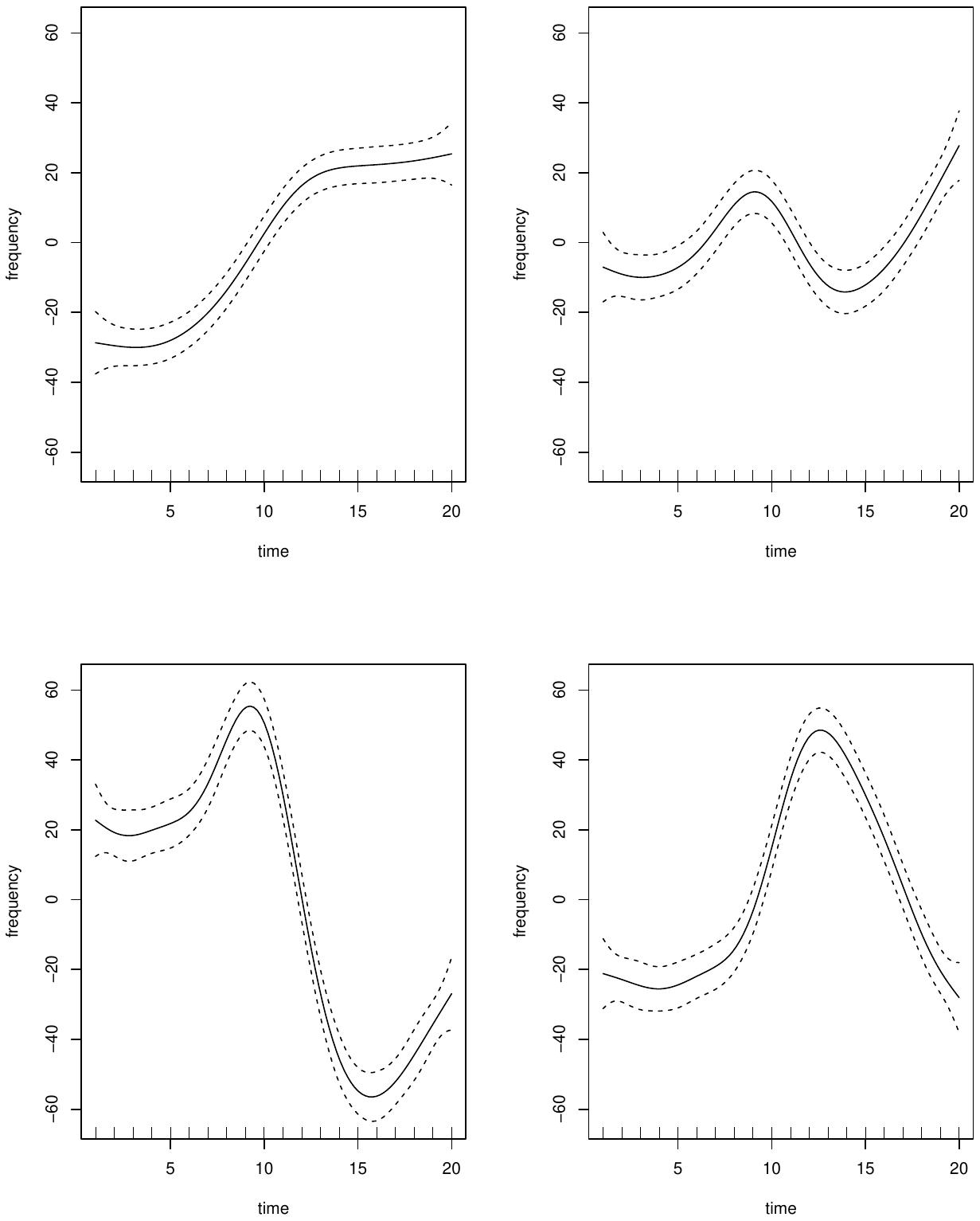}\\ \includegraphics[width=12cm, height=5.5cm]{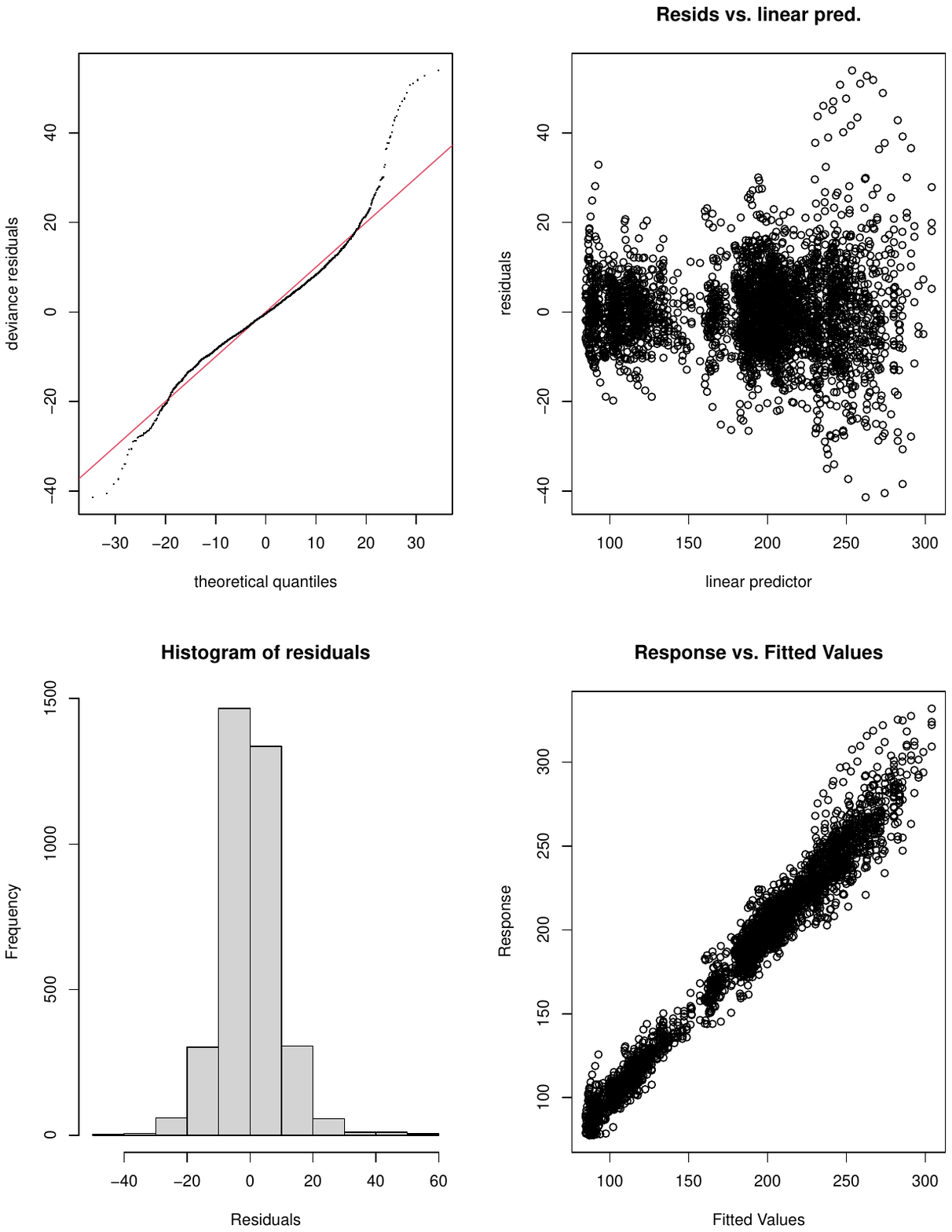}
\end{tabular}
\caption{Average of f0 curves across repetitions and speakers of T2Tx (above) and GAM model check for T2Tx f0 contours (under).}
    \label{fig:t2tx}
\end{table}

\newpage 

\subsection{GAM model for T3Tx combination}
\vspace{-0.5cm}
\begin{table}[h!]
  \label{table:gamT3Tx}
  \centering
  \begin{tabular}{|ll||ll|}
    \hline
 \textbf{Parametric coefficients} & \textbf{p-value} &\textbf{Smooth terms} &  
                                                             \textbf{p-value}\\
    \hline
    (Intercept) &  $<0.01^{***}$ & s(time): no CL T3.T1  &  $0.01^{***}$\\
    CL T3.T1 &  $0.01^{*}$ & s(time): CL T3.T1 &  $0.01^{***}$\\
    no CL T3.T2  & $0.04^{*}$ & s(time): no CL T3.T2  & $0.01^{***}$\\
    CL T3.T2 & $<0.01^{***}$ & s(time): CL T3.T2   & $0.01^{***}$\\
    no CL T3.T3  & $<0.01^{***}$ & s(time): no CL T3.T3   & $0.01^{***}$\\
    CL T3.T3  & $<0.01^{***}$ & s(time): CL T3.T3  & $0.01^{***}$\\
    no CL T3.T4  & $<0.09$ & s(time): no CL T3.T4  & $<0.01^{***}$\\
    CL T3.T4  & $0.01^{*}$ & s(time): CL T3.T4  & $0.01^{***}$\\
    && s(time, subject interaction)  & $<0.01^{***}$\\
    \hline
  \end{tabular}
\caption{GAM output to model anticipatory tonal coarticulation effect in T3Tx combination.}
\end{table}

\begin{table}[h!]
\begin{tabular}{l}
    \includegraphics[width=12cm, height=5.5cm]{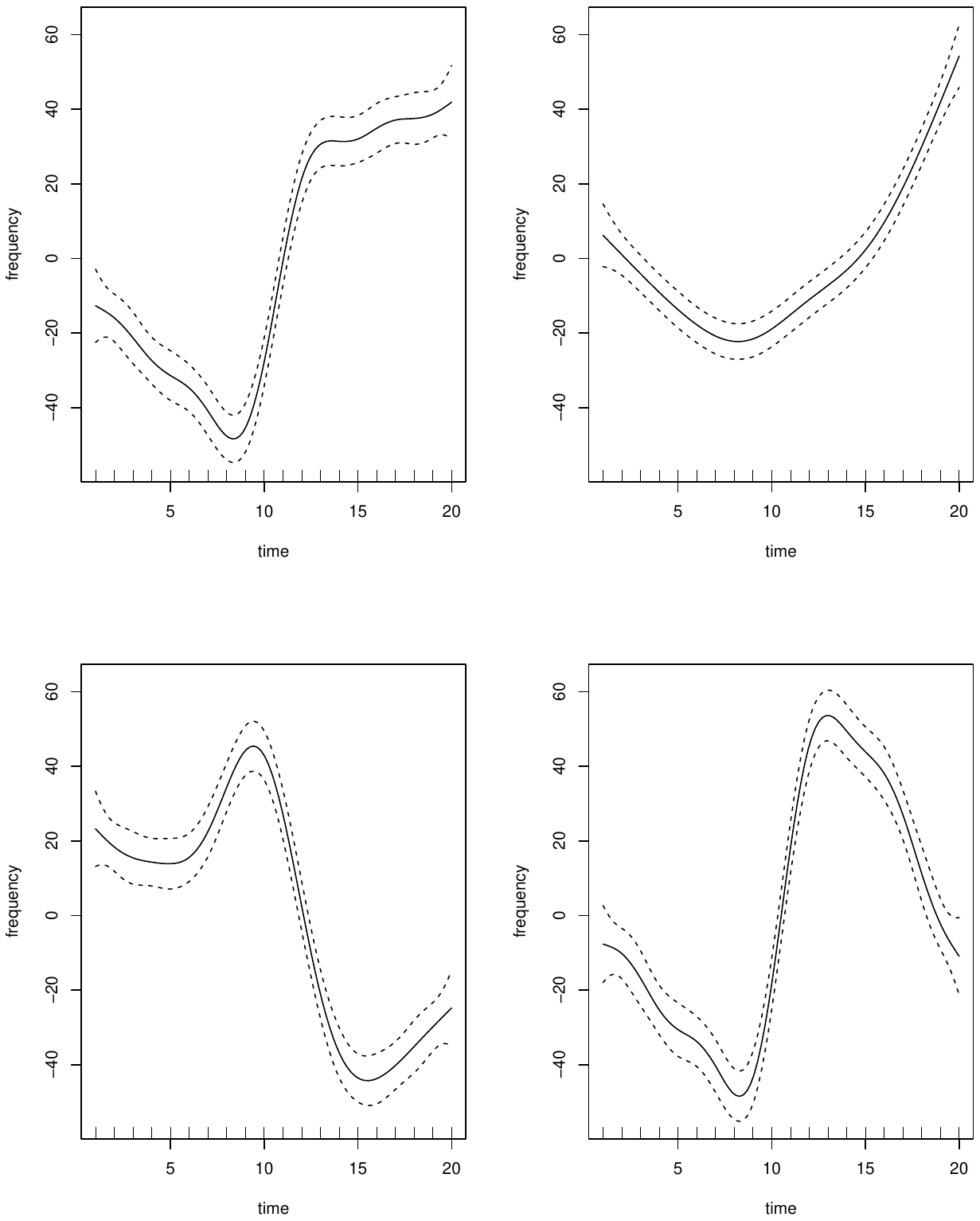}\\ \includegraphics[width=12cm, height=5.5cm]{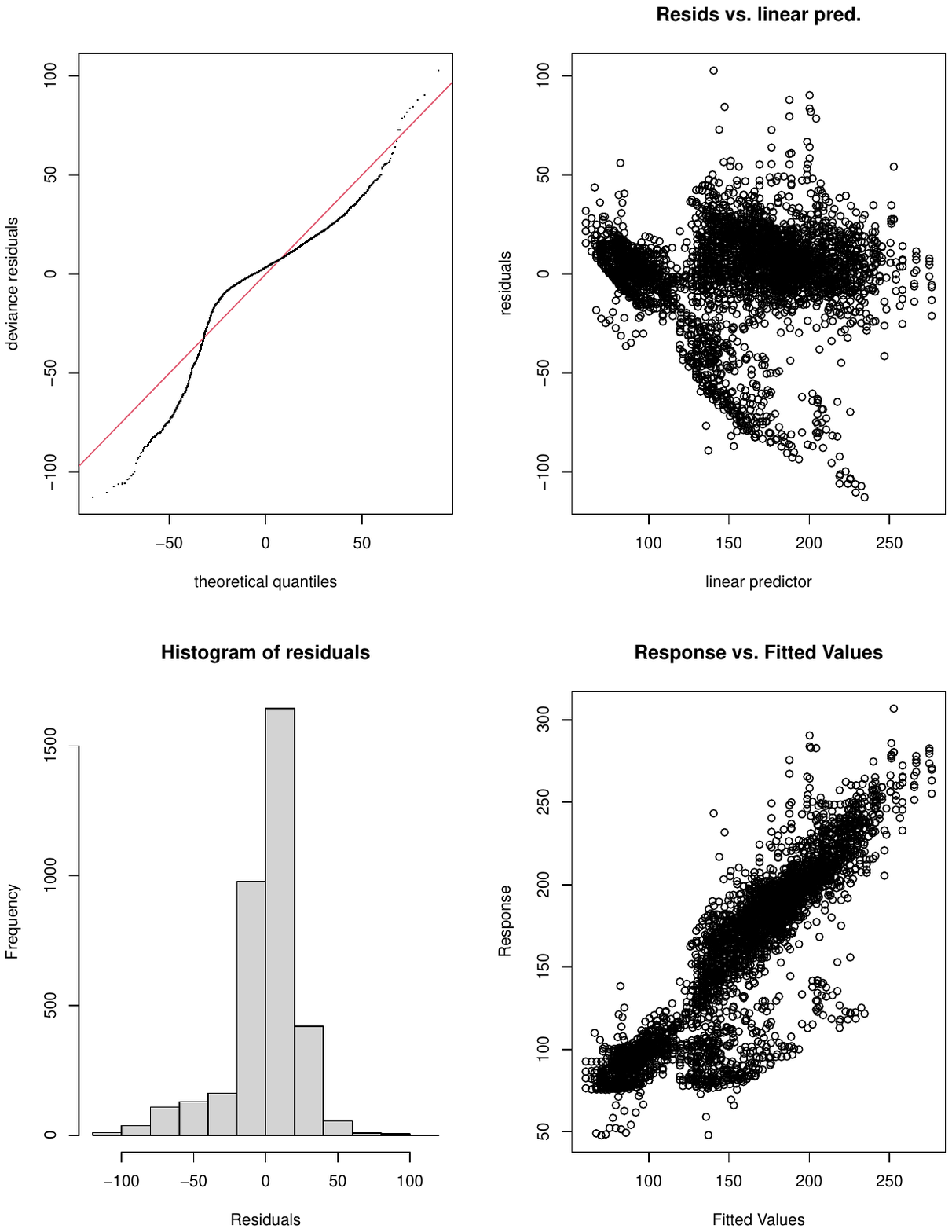}
\end{tabular}
\caption{Average of f0 curves across repetitions and speakers of T3Tx (above) and GAM model check for T3Tx f0 contours (under).}
    \label{fig:t3tx}
\end{table}

\newpage

\subsection{GAMM model for T4Tx combination}
\vspace{-0.5cm}
\begin{table}[h!]
  \label{table:gamT4Tx}
  \centering
  \begin{tabular}{|ll||ll|}
    \hline
 \textbf{Parametric coefficients} & \textbf{p-value} &\textbf{Smooth terms} &  
                                                             \textbf{p-value}\\
    \hline
    (Intercept) &  $<0.01^{***}$ & s(time): no CL T4.T1  &  $0.01^{***}$\\
    CL T4.T1 &  $0.28$ & s(time): CL T4.T1 &  $0.01^{***}$\\
    no CL T4.T2  & $<0.01^{***}$ & s(time): no CL T4.T2  & $0.01^{***}$\\
    CL T4.T2 & $0.02^{*}$ & s(time): CL T4.T2   & $0.01^{***}$\\
    no CL T4.T3  & $<0.01^{***}$ & s(time): no CL T4.T3   & $0.01^{***}$\\
    CL T4.T3  & $<0.01^{***}$ & s(time): CL T4.T3  & $0.01^{***}$\\
    no CL T4.T4  & $<0.09$ & s(time): no CL T4.T4  & $<0.01^{***}$\\
    CL T4.T4  & $0.04^{*}$ & s(time): CL T4.T4  & $0.01^{***}$\\
    && s(time, subject interaction)  & $<0.01^{***}$\\
    \hline
  \end{tabular}
\caption{GAM output to model anticipatory tonal coarticulation effect in T4Tx combination.}
\end{table}

\begin{table}[h!]
\begin{tabular}{l}
    \includegraphics[width=12cm, height=5.5cm]{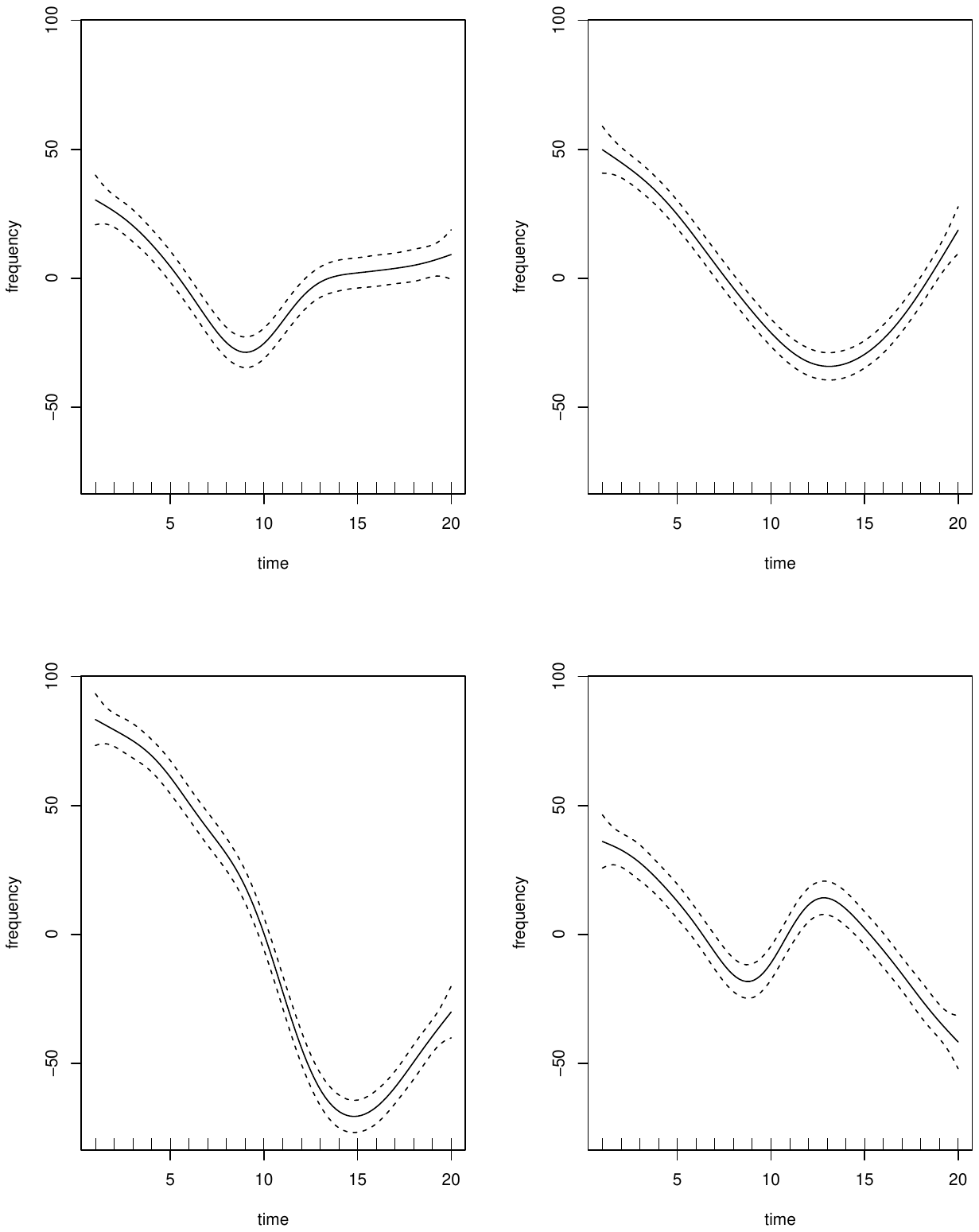}\\ \includegraphics[width=12cm, height=5.5cm]{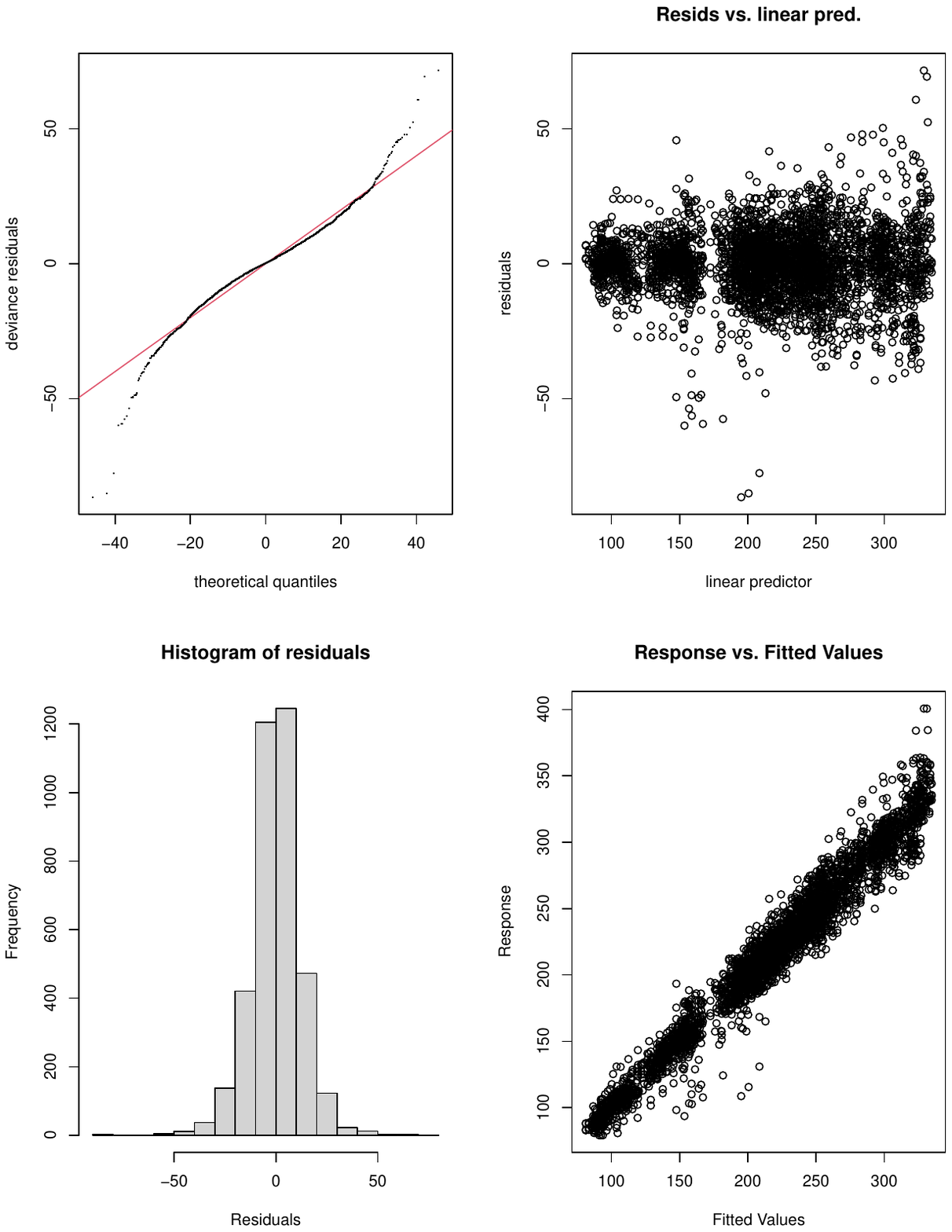}
\end{tabular}
\caption{Average of f0 curves across repetitions and speakers of T4Tx (above) and GAM model check for T4Tx f0 contours (under).}
    \label{fig:t4tx}
\end{table}

\end{document}